\documentstyle[12pt]{article}
\input epsf
\begin{document}

\def\question#1{{{\marginpar{\small \sc #1}}}}
\newcommand{\qcd'}{$\tilde{\rm QCD}$}
\newcommand{\eq}{\begin{equation}}
\newcommand{\en}{\end{equation}}
\newcommand{\bino}{\tilde{b}}
\newcommand{\tsquark}{\tilde{t}}
\newcommand{\gluino}{\tilde{g}}
\newcommand{\photino}{\tilde{\gamma}}
\newcommand{\wino}{\tilde{w}}
\newcommand{\mtilde}{\tilde{m}}
\newcommand{\higgsino}{\tilde{h}}
\newcommand{\gsi}{\,\raisebox{-0.13cm}{$\stackrel{\textstyle>}
{\textstyle\sim}$}\,}
\newcommand{\lsi}{\,\raisebox{-0.13cm}{$\stackrel{\textstyle<}
{\textstyle\sim}$}\,}

\rightline{RU-95-17}
\rightline{hep-ph/9504295}
\rightline{April 12, 1995}
\baselineskip=18pt
\vskip 0.4in
\begin{center}
{\bf \LARGE Phenomenology of Light Gauginos}\\
\vspace*{0.4in}
{\large Glennys R. Farrar}\footnote{Research supported in part by
NSF-PHY-91-21039} \\
\vspace{.1in}
{\it Department of Physics and Astronomy \\ Rutgers University,
Piscataway, NJ 08855, USA}\\
\end{center}
\vspace*{0.2in}
\vskip  0.1in

{\bf Abstract:} I advocate the virtues of a very economical version of
the minimal supersymmetric standard model which avoids cosmological
problems often encountered in dynamical SUSY-breaking and solves the
SUSY-CP problem.  Imposing $m_Z=91$ GeV and $m_t \sim 175$ GeV implies
that scalar masses are generally $100-200$ GeV.  The gluino and
photino are massless at tree level.  At 1-loop, the gluino mass is
predicted to be in the range $m_{\gluino}: 100 - 600$ MeV and the
photino mass can be estimated to be $m_{\photino}: 100 - 1400$ MeV.
New hadrons with mass $\sim 1 \frac{1}{2}$ GeV are predicted and
described.  The ``extra'' flavor singlet pseudoscalar observed in two
experiments in the $\iota(1440)$ region, if confirmed, is naturally
interpreted as the state which gets its mass via the QCD anomaly.  Its
superpartner, a gluon-gluino bound state, generally has a lifetime
longer than $5~10^{-11}$ sec and would not have shown up in existing
searches.  Search strategies and other consequences of the scenario
are discussed.
\thispagestyle{empty}
\newpage
\addtocounter{page}{-1}
\newpage

The customary approach to studying the phenomenological implications
of supersymmetry has been to assume that the ``low energy'' effective
Lagrangian below the SUSY-breaking scale, $M_{SUSY}$, contains all
possible renormalizable operators, including in principle all possible
soft supersymmetry breaking terms, consistent with the gauge
symmetries and possibly some global and discrete symmetries.  Some
models of SUSY-breaking naturally lead to relations among the SUSY-breaking
parameters at the scale $M_{SUSY}$ so that the minimal supersymmetric
standard model (MSSM) requires specification of 6-8 parameters beyond
the gauge and Yukawa couplings already determined in the MSM: $tan
\beta \equiv \frac{v_U}{v_D}$, the ratio of the two Higgs vevs, $\mu$,
the coefficient of the SUSY-invariant coupling between higgsinos,
$M_0$, a universal SUSY-breaking scalar mass, $m_{12}^2$, the
SUSY-breaking mixing in the mass-squared matrix of the Higgs scalars
(aka $\mu B$ or $\mu M_0 B$ in alternate notations), $M_{1,2,3}$, the
SUSY-breaking gaugino masses (proportional to one another if the MSSM
is embedded in a GUT), and $A$, the coefficient of SUSY-breaking terms
obtained by replacing the fermions in the MSM Yukawa terms by their
superpartners.  To obtain predictions for the actual superparticle
spectrum in terms of these basic parameters, the renormalization group
equations for masses, mixings and couplings are evolved from the scale
$M_{SUSY}$ to the scale $M_{Z^0}$ where on account of different RG
running and flavor dependent couplings, the various scalars and
fermions have quite different masses.  A particularly attractive
aspect of this approach is that for the heavy top quark which is found
in nature\cite{cdf:top,D0:top}, the mass-squared of a combination of
Higgs fields becomes negative at low energy and the electroweak
symmetry is spontaneously broken\cite{a-gpw,ir}, with $m_Z$ a function
of $A,~ M_0$ and other parameters of the theory. In this conventional
treatment of the MSSM, the lightest squark mass is constrained by
experiment to be greater than 126 GeV and the gluino mass to be
greater than 141 GeV\cite{cdf:gluinolim2}.

I will argue here that a more restrictive form of low energy SUSY
breaking is actually more plausible, one without dimension-3
operators.  We shall see that the remaining parameters of the theory
are well-constrained when electroweak symmetry breaking is demanded, and
that the resultant model (MSSM') is both extremely predictive and
consistent with laboratory and cosmological observations.  If this is
the correct structure of the low energy world, there will be many
consequences which can be discovered and investigated before the
construction of the LHC.  Some of these are discussed below.

There are at least two good reasons for dispensing with dimension-3
SUSY breaking operators in the low energy theory.  In the most
attractive SUSY-breaking scenario, hidden sector dynamical SUSY
breaking, such operators make negligible contribution unless there is
one or more {\it gauge singlet} field in the hidden sector whose
auxilliary field gets a large vev.  However it was shown in
ref. \cite{bkn} that SUSY breaking with hidden sector gauge singlets
lead to particles with masses in the 100 GeV - 1 TeV region which are
in conflict with cosmology, in particular causing late-time entropy
production which is incompatible with primordial nucleosynthesis.
Besides avoiding the problems associated with singlet fields, not
having dimension-3 SUSY breaking is attractive because it solves the
SUSY CP problem\footnote{In this scenario, the only phases other than
those associated with the strong CP problem (the phases in the quark
mass matrix and $\theta$ parameter) which can be present in the theory
at the scale $M_{SUSY}$ appear in the parameters $\mu$ and
$m_{12}^2$. However a combination of an $R$-transformation and U(1)
transformations on the Higgs superfields allows these phases to be
removed.  Any phase which is introduced thereby into the Yukawa terms
in the superpotential can be removed by chiral transformations on the
quark superfields, merely changing the phases which contribute to the
strong CP problem (which must be solved by some other mechanism).
Since the gauge-kinetic terms are not affected by U(1) and $R$
transformations, the preceding manipulations do not introduce phases
in interactions involving gauginos.  I thank Scott Thomas for pointing
out how use of the $R$ transformation simplifies the proof.}.

If there are no dimension-3 SUSY-breaking operators, $A$ and
$M_{1,2,3}$ are zero, and the gluino and lightest neutralino are
massless in tree approximation.  They get masses at one loop from
virtual top-stop pairs, and, for the neutralinos, from ``electroweak''
loops involving wino/higgsino-Higgs/vector boson
pairs\cite{bgm,pierce_papa,f:96}.  The size of these corrections were
estimated, for various values of $M_0, ~\mu$, and $ tan \beta$ in ref.
\cite{f:96}.  There, it was determined that in order to insure that the
chargino mass is greater than its LEP lower bound of about 45
GeV, $\mu$ must either be less than 100 GeV (and $tan \beta \lsi 2$) or
greater than several TeV.  Here I will also demand that the
electroweak symmetry breaking produces the observed $m_Z$ for $m_{t}
\sim 175$ GeV\cite{cdf:top}.  This is not possible in the large $\mu$
region, so I will consider only $\mu \lsi 100$ GeV.  In addition, from
Fig. 6 of ref. \cite{a-gpw} one sees that $M_0$, the SUSY-breaking
scalar mass, must be $\sim 100 - 300 $ GeV, with 150 GeV being the
favored value.  From Figs. 4 and 5 of ref. \cite{f:96} this gives
$m_{\gluino} \sim 100-600$ and $m_{\photino} \sim 100-900$ MeV. Since
the electroweak loop was treated in ref. \cite{f:96} with an
approximation which is valid when $M_0$ or $\mu$ is $>> m_Z$, their
results for the photino mass are only indicative of the range to be
expected.  Until a more precise calculation is available, we attach a
$\sim$factor-of-two uncertainty to the electroweak loop, and consider
the enlarged photino mass range $100 - 1400$ MeV.

The purpose of this Letter is to investigate the most essential
aspects of the phenomenology of this theory, having restricted quite
substantially the allowed ranges of parameters.  The primary issues to
be discussed are:
\begin{enumerate}

\item  Predicted mass and lifetime of the lightest $R$-meson, the $g
\gluino$ bound state denoted $R^0$.

\item  Predicted mass of the flavor singlet pseudoscalar which gets
its mass via the anomaly (the ``extra'' pseudoscalar corresponding to
the $ \gluino \gluino$ ground state degree of freedom).

\item  Identity of the flavor singlet pseudogoldstone boson resulting
from the spontaneous breaking of the extra chiral symmetry associated
with the light gluino.

\item  The flavor-singlet $R$-baryon composed of $uds\gluino$, called
$S^0$.

\item  Production rates and detection strategies for the new $R$-hadrons.

\end{enumerate}

The $R^0$ lifetime is the most important item in this list.  For many
years the famous UA1 figure\cite{ua1} and its decendants, showing the
allowed regions of the gluino-squark mass plane, has widely been
accepted as excluding all but certain small ``windows'' for low gluino
mass.  However this figure was constructed under the {\it assumption}
that the gluino lifetime is short enough that missing energy and beam
dump experiments are sensitive to it.  As emphasized in
ref. \cite{f:55}, $R$-hadrons produced in the target or beam dump
degrade in energy very rapidly due to their strong interactions.
However the photino is supposed to reinteract in the detector
downstream of the beam dump or carry off appreciable missing energy.
Only when it is emitted before the $R$-hadron interacts, will it
typically have enough energy to be recognized.  As discussed in
connection with a particular experiment in ref. \cite{f:55}, and more
generally in ref. \cite{f:95}, if the $R^0$ lifetime is longer than
$\sim 5~10^{-11}$ sec this criterion is not met and beam dump and
missing energy experiments become ``blind'' to light gluinos. Thus the
commonly accepted notion of various tiny ``windows'' for light
gluinos, is simply wrong for the case that $R^0$'s have lifetimes
longer than $\sim 5~10^{-11}$ sec.\footnote{The cutoff in sensitivity
as a function of $\tau_{\photino}$, the characteristic timescale for
photino emission, depends on the experiment; $5~10^{-11}$ is just a
useful ballpark demarcation.  Unfortunately, the acceptance of these
experiments has not generally been reported in experimental terms, as
a function of $\tau_{\photino}$, $\sigma_{\gluino}$ and
$\sigma_{\photino}$ (gluino production and photino rescattering cross
sections).  Instead results have been reported in terms of excluded
regions in the $m_{\gluino},~m_{sq}$ plane, implicitly assuming that
$\tau_{\photino}$ is small enough for the gluino energy not to have
been degraded before photino emission.  Had the former information
been reported, it would be possible to make a more quantitative
statement on the allowed $R^0$ lifetime: since gluinos are produced
primarily in $R$-mesons which decay quickly to an $R^0$\cite{f:95},
$\tau_{R^0} \approx \tau_{\photino}$.}  Furthermore, the UA1 analysis
accepted at face value a number of experiments searching for low-mass
gluinos which were analyzed using perturbative QCD predictions which
fail to make the important distinction between the current mass of the
gluino and its constituent mass or the mass of the hadrons containing
it.  As will be clear after we find the $R^0$ mass in the massless
gluino limit, this distinction is even more important than in the case
of quarks.  Ignoring it leads in some cases to a serious overestimate
of the expected production rate, and thus an exagerated view of the
experimental sensitivity.

\begin{figure}
\epsfxsize=\hsize
\epsffile{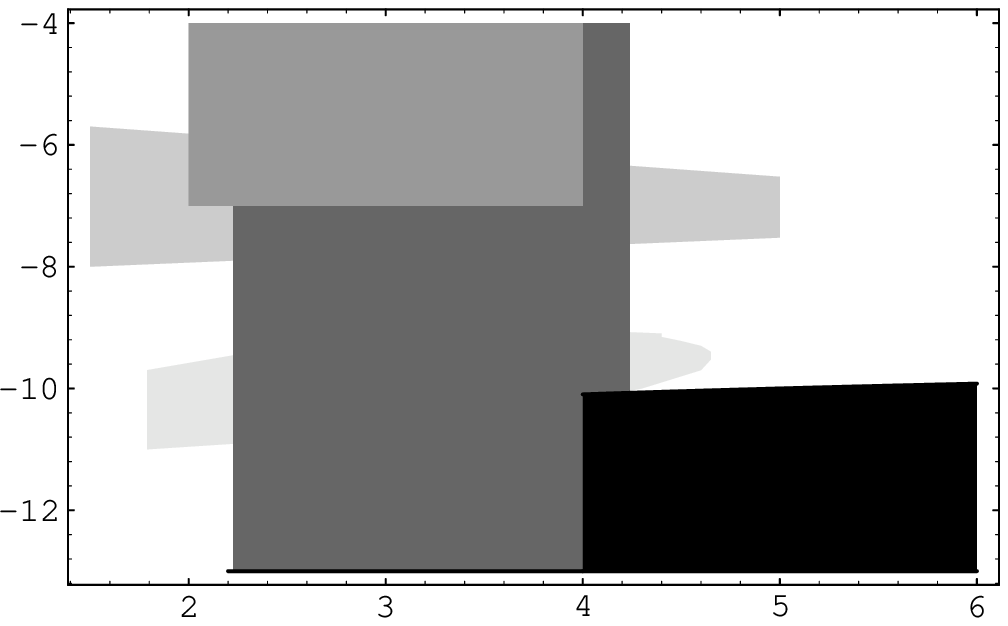}

\caption{Figure showing the experimentally excluded regions of $m(R^0)$ and
$\tau_{\tilde{g}}$.  Horizontal axis is $m(R^0)$ in GeV beginning at
1.5 GeV; vertical axis is $Log_{10}$ of the lifetime in sec.  A
massless gluino would lead to $m(R^0) \sim 1.4 \pm .4$ GeV.  ARGUS and
Bernstein et al give the lightest and next-to-lightest regions (lower
and upper elongated shapes), respectively.  CUSB gives the
next-to-darkest block; its excluded region extends over all
lifetimes. Gustafson et al gives the smaller (mid-darkness) block in
the upper portion of the figure; it extends to infinite lifetime.  The
UA1 experiment the darkest block in the lower right corner; it extends
to higher masses and shorter lifetimes not shown on the figure.}
\label{ltgl}
\end{figure}

In ref. \cite{f:95} I reported the result of a comprehensive study of
relevant experiments, including all those used in the UA1 analysis.
The problems mentioned in the previous paragraph and the modifications
they imply in the analysis of the excluded regions are discussed in
detail in ref. \cite{f:95}.  Much of the low-mass region is only
excluded when beam-dump experiments are applicable.  Instead one finds
(see Fig. \ref{ltgl}, reproduced from ref. \cite{f:95}) that an $R^0$
in the mass range ($\sim 1.1-1.5$ GeV) is consistent with the present
experimental situation for any lifetime $\gsi 5~10^{-11}$ s. The mass
range $\sim 1.5 - 1.8$ GeV is allowed for lifetimes between $\sim
5~ 10^{-11} - 10^{-8}$s or $\gsi 2 \times 10^{-6}$s.  In the restrictive
scenario under discussion, $M_0$ is rather well determined so we will
try below to estimate the $R^0$ lifetime.

In order to estimate the $R^0$ lifetime, we need its mass.
Fortunately, it can be quite well determined from existing lattice QCD
calculations, as follows\cite{f:95}. If the gluino were massless and
there were no quarks in the theory (let us call this theory sQCD),
SUSY would be unbroken and the $R^0$ would be in a degenerate
supermultiplet with the $0^{++}$ glueball, $G$, and a $0^{-+}$ state I
shall denote $ \tilde{\eta}$, which can be thought of as a $\gluino
\gluino$ bound state\footnote{It is convenient to think of the states
in terms of their ``valence'' constituents but of course each carries
a ``sea'' so, e.g., the glueball may be better described as a coherent
state of many soft gluons than as a state of two gluons.  Knowledge of
these aspects of the states is not needed for estimation of their
masses.}. To the extent that quenched approximation is accurate for
sQCD\footnote{The 1-loop beta function is the same for sQCD as for
ordinary QCD with 3 light quarks, so the accuracy estimate for
quenched approximation in ordinary QCD, $10-15\%$, is applicable
here.}, the mass of the physical $R^0$ in the continuum limit of this
theory would be the same as the mass of the $0^{++}$ glueball, which
has been measured in quenched lattice QCD to be $1440 \pm 110$
MeV\cite{weingarten:glueballs}.  Including errors associated with
unquenching the quarks and gluinos, the $G,~R^0$, and $\tilde{\eta}$
masses were estimated in ref. \cite{f:95} to be $1440 \pm 375$
MeV\footnote{The uncertainty coming from unquenching both light quarks
and gluinos was taken there to be 25\%, however to the extent that the
estimate of the quenching error for ordinary QCD is obtained by
comparing lattice results with the hadron spectrum, that will already
include the effects of gluinos if they are present in nature, and the
$\pm375$ MeV uncertainty should be replaced by $\pm 240$ MeV.}.
Mixing with other flavor singlet pseudoscalars can shift the
$\tilde{\eta}$ somewhat.  For gluino masses small compared to the
``confinement mass'' of $\sim 1 \frac{1}{2}$ GeV, one would expect the
$R^0$ and $\tilde{\eta}$ masses to be insensitive to the gluino mass.
Thus in the absence of a dedicated lattice gauge theory calculation of
the masses of these particles, we can adopt the estimate, $1.4 \pm
0.4$ GeV for all these states, $G,~\tilde{\eta}$, and $R^0$.

Having a reliable mass estimate is an important component in
determining the phenomenology of a light gluino.  Bag model
estimates\cite{chanowitz_sharpe} were significantly lower than this,
as they were for the glueball spectrum in comparison to lattice gauge
theory.  The fact that the $R^0$ mass is so much greater than even the
vector meson masses, means that its production at low energies will be
substantially kinematically suppressed compared to naive pQCD
estimates.  This fact, which was neglected in the UA1 analysis, was
incorporated in the analysis of ref. \cite{f:95}.

Note that in sQCD, which is identical to ordinary QCD in quenched
approximation, the $\tilde{\eta}$ with mass $\sim1.44$ GeV is the
pseudoscalar which gets its mass from the anomaly.  Thus in QCD with
light gluinos (QCD') the particle which gets its mass from the anomaly
is too heavy to be the $\eta'$.  Instead, the $\eta'$ should be
identified with the pseudogoldstone boson associated with the
spontaneous breaking of the non-anomalous chiral U(1)\footnote{Formed
from the usual chiral U(1) of the light quarks and the chiral
R-symmetry of the gluinos\cite{f:95}.} by the formation of $q \bar{q}$
and $\gluino \gluino$ condensates, $<\bar{q} q>$ and
$<\bar{\lambda}\lambda>$.  Using the usual PCAC and current algebra
techniques, in ref. \cite{f:95} I obtained the relationship between
masses and condensates necessary to produce the correct $\eta'$ mass
(ignoring mixing): $m_{\gluino} <\bar{\lambda}\lambda>~ \sim 10~ m_s
<s \bar{s}>$.  The required gluino condensate is reasonable, for
$m_{\gluino} \gsi 100$ MeV\footnote{Ensuring $m_{\gluino} \gsi 100$
requires $M_0 \lsi 300$ GeV\cite{f:96}.  This is consistent with the
values indicated by electroweak symmetry breaking for $m_t=175$
GeV.}. In a more refined discussion, the physical $\eta'$ would be
treated as a superposition of the pseudo-goldstone boson and the
orthogonal state which gets its mass from the anomaly.  The important
point, independent of details of the mixing, is that this scenario
{\it predicts} the existance of a flavor singlet pseudoscalar meson in
addition to the $\eta'$ which is not a part of the conventional QCD
spectrum of quark mesons and glueballs, whose mass should be $\sim 1.4
\pm 0.4$ GeV.  I have not yet identified any clear test for the
prediction that the $\eta'$ is mainly a pseudogoldstone boson and
probably contains a $\sim 30 \%$ $\gluino \gluino$ component, since
model independent predictions concerning the $\eta'$ are for ratios in
which the gluino component plays no role.\footnote{Chiral perturbation
theory implies characteristic relations between various physical
quantities involving pseudogoldstone bosons.  Whether it can be used
here, given the large mass of the $\eta'$, is under investigation.
GRF and M. Luty, in preparation.}

Having in hand an estimate of the $R^0$ mass and photino mass, we now
return to determining the $R^0$ lifetime.  Making an absolute estimate
of the lifetime of a light hadron is always problematic.  Although the
relevant short distance operators can be accurately fixed in terms of
the parameters of the Lagrangian, hadronic matrix elements are
difficult to determine.  It is particularly tricky for the $R^0$ in
this scenario because the photino mass is larger than the current
gluino mass and, since $m_{\photino} \sim \frac{1}{2} m_{R^0}$, the
decay is highly suppressed even using a constituent mass for the
gluino.  The decay rate of a free gluino into a photino and massless
$u \bar{u}$ and $d \bar{d}$ pairs is known\cite{hk:taugluino}:
\eq
\Gamma_0(m_{\gluino},m_{\photino}) = \frac{\alpha \alpha_s
m^5_{\gluino}}{48 \pi M_{sq}^4} \frac{5}{9}
f(\frac{m_{\photino}}{m_{\gluino}}),
\label{GammaR0}
\en
taking $M_{sq}$ to be a common up and down squark mass.  The function
$f(y)=[(1-y^2)(1 + 2y - 7y^2 + 20y^3 - 7 y^4 + 2 y^5 + y^6) +
24y^3(1 - y + y^2)log(y)]$ contains the phase space suppression which
is important when the photino is massive.  The problem is to take
into account how interactions with the gluon and ``sea'' inside the
$R^0$ ``loans'' mass to the gluino.  If this effect is ignored one
would find the $R^0$ to be absolutely stable except in the upper
portion of its estimated mass range.

\begin{figure}
\epsfxsize=\hsize
\epsffile{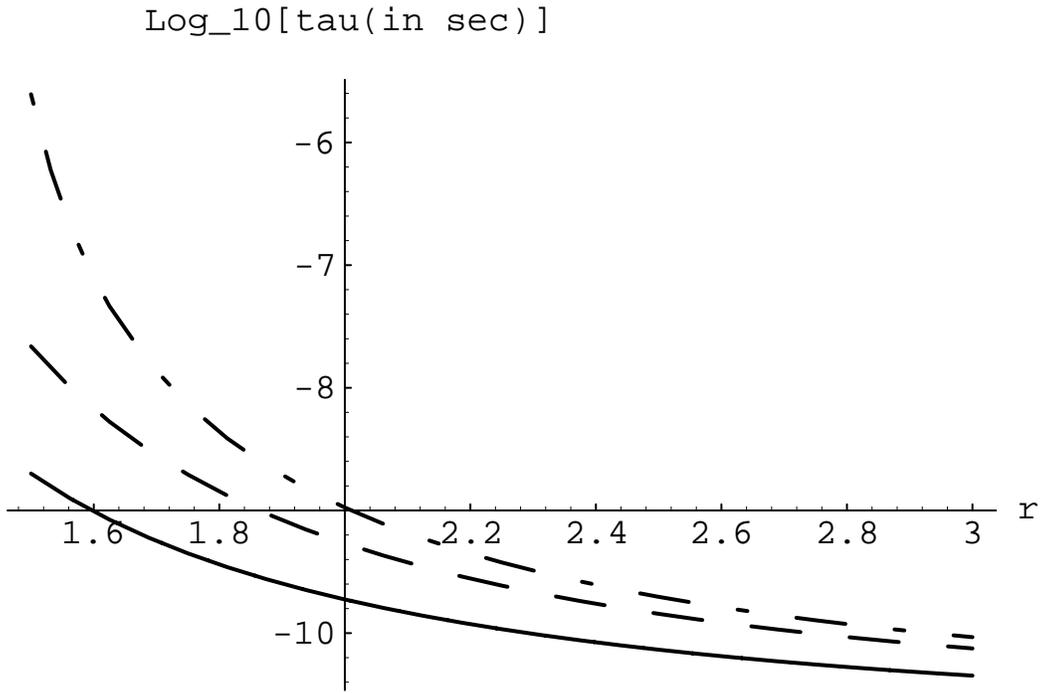}

\caption{$R^0$ lifetime in a crude model for three different gluon
distribution functions described in the text (solid: $F_{ur}$,
dashed: $F_{10}$, dot-dashed: $F_{nr}$) as a function of $r\equiv
\frac{m(R^0)}{m_{\photino}}$, with $m(R^0) = 1.5$ GeV and $M_{sq}=150$
GeV.}
\label{tau_wfn}
\end{figure}

A method of estimating the {\it
maximal} effect of such a ``loan'', and thus a lower limit on the
$R^0$ lifetime, can be obtained by elaborating a suggestion of refs.
\cite{accmm,franco}.  The basic idea is to think of the hadron as a
bare massless parton (in this case a gluon) carrying momentum fraction
$x$ and a remainder (here, the gluino) having an effective mass $M
\sqrt{1-x}$, where $M$ is the mass of the decaying hadron, here the
$R^0$.  Then the structure function, giving the probability
distribution of partons of fraction $x$, also gives the distribution
of effective masses for the remainder (here, the gluino).  Summing the
decay rate for gluinos of effective mass $m(R^0)\sqrt{1-x}$ over the
probability distribution for the gluino to have this effective mass,
leads to a crude estimate or upper bound on the rate:
\eq
\Gamma(m(R^0),z) = \Gamma_0(m(R^0),0) \int^{1-z^2}_0
(1-x)^{\frac{5}{2}} F(x) dx f(z/\sqrt{1-x}),
\en
where $z=\frac{m_{\photino}}{m(R^0)}$.  The distribution function of
the gluon in the $R^0$ is unknown, but can be bracketed with extreme
cases: the non-relativistic $F_{nr}(x) = \delta(x-\frac{1}{2})$
and the ultrarelativistic $F_{ur}(x)= 6x(1-x)$.  The normalizations
are chosen so that half the $R^0$'s momentum is carried by gluons.
Figure \ref{tau_wfn} shows the $R^0$ lifetime produced by this model,
for $M_{sq} = 150$ GeV and $m(R^0) = 1.5$ GeV, for these two structure
functions, and also for the intermediate choice $F_{10}(x) = N_{10}
x^{10}(1-x)^{10}$, as a function of $r\equiv
z^{-1}=\frac{m(R^0)}{m_{\photino}}$.  Results for any $R^0$ and
squark mass can be found from this figure using the scaling behavior
$\Gamma(m(R^0),M_{sq},z) \sim m(R^0)^{5} M_{sq}^{-4} g(z)$, as long as
it is legitimate to ignore the mass of the remnant hadronic
system, say a pion.

The decay rates produced in this model can be considered upper limits
on the actual decay rate, because the model in some sense maximizes
the ``loan'' in dynamical mass which can be made by the gluons to the
gluino.  For kaon semileptonic decay (where the $K_{\mu 3}$ mode would
be strongly suppressed or excluded in the absence of similar effects,
since the strange quark current mass is of the same order as the muon
mass) this model gives the correct ratio between $K_{\mu 3}$ and $K_{e
3}$ rates, and rates 2-4 times larger than observed: overestimating
the rate as we anticipated, but not by a terribly large factor.
Although a large range of uncertainty should be attached to the $R^0$
lifetime estimated this way, these results are still useful because
they give lower bounds on the lifetime.  We can see that the beam dump
experiments\cite{charm,ball,bebc,akesson} are unlikely to have been
sensitive to the $R^0$.  Even with the ultrarelativistic wavefunction
which gives the shortest lifetime estimate, for most of the parameter
space of interest the lifetime is not short enough for the $R^0$ to
have decayed before its energy is degraded by interactions in the
dump, which requires $\tau(R^0) \lsi 5 \times 10^{-11}$\cite{f:95}.

There is another interesting light $R$-hadron besides the $R^0$,
namely the flavor singlet scalar baryon $uds \gluino$ denoted $S^0$.
In view of the very strong hyperfine attraction among the
quarks\cite{f:51}, this state may be similar in mass to the
$R^0$.\footnote{It is amusing that the (spin 1/2) baryon spectrum
contains an anomalous state, the $\Lambda(1405)$, which this
discussion suggests is likely to be a $uds$-gluon flavor singlet
cryptoexotic baryon.  The validity of this suggestion for identity of
the $\Lambda(1405)$ is of course independent of the existance of light
gluinos.}  If its mass is $\sim 1 \frac{1}{2}$ GeV, it will be
extremely long lived, or even stable if $m(S^0) - m(p) - m(e^-) <
m_{\photino}$.  Even if decay to a photino and nucleon is
kinematically allowed, the decay rate will be very small since it
requires a flavor-changing-neutral transition as well as an
electromagnetic interaction.  If the $S^0$ does not bind to nuclei,
its being absolutely stable is not experimentally
excluded\cite{f:51,f:95}.  There is not a first-principles
understanding of the  intermediate-range nuclear force, so that it is
not possible to decide with certainty whether the $S^0$ will bind to
nuclei.  However the two-pion-exchange force, which is attractive
between nucleons but insufficient to explain their binding\footnote{So
that an ad hoc phenomenological $\sigma$ exchange is postulated, which
does not however correspond to any observed particle. I am indebted to
R. Amado for discussions of this point.}, is repulsive in this
case\cite{f:95} because the mass of the intermediate $R_{\Lambda}$ or
$R_{\Sigma}$ is much larger than that of the $S^0$.\footnote{If its
decay to a nucleon is kinematically forbidden, the $S^0$ would be
absolutely stable unless both $R$-parity and baryon-number
conservation are violated.  In some models of $R$-parity
violation\cite{masiero_valle}, $R$-parity is violated in association
with lepton number violation, but baryon number is conserved.  In this
case the photino would be unstable (e.g., $\photino \rightarrow \nu
\gamma$) but the $S^0$ stable.  Stable relic $S^0$'s could make up
part of the missing mass in our galaxy, but since they have strong
interactions they would clump too much to account for the bulk of the
missing dark matter of the universe.  I thank E. Kolb for discussions
on this matter.}  For further discussion of the $S^0$ and other
$R$-hadrons see refs. \cite{f:51} and \cite{f:95}.

We have seen above that existing searches for gluinos and $R$-hadrons
do not exclude this scenario, but it is also commonly claimed that light
photinos are excluded.  However those arguments do not apply to the case
at hand. First of all, since gaugino masses come from radiative
corrections, limits relying on GUT tree-level relations between
gaugino masses do not apply.  Furthermore, the lightest and
next-to-lightest neutralinos (called in general $\chi^0_1$ and
$\chi^0_2$) are not produced in $Z$ decays with sufficient rate to be
observed at LEP, because in this scenario the $\chi^0_1$ is extremely
close to being pure photino\cite{f:96}.  It contains so little
higgsino that $Z^0 \rightarrow \chi_1^0 \chi_1^0$ and $Z^0 \rightarrow
\chi_1^0 \chi_2^0$ are suppressed compared to the conventional
scenario with tree-level gaugino masses, while the mass of the
$\chi^0_2$ is high enough that its pair production is too small to be
important for most of parameter space.

It has also been claimed that a stable photino with mass less than 10
GeV is excluded because it would produce too large a relic abundance,
``overclosing'' the universe.  These calculations assumed that
self annihilation was the only important mechanism for keeping
photinos in thermal equilibrium, leading to freeze out at a
temperature $T \sim \frac{1}{14} m_{\photino}$, below which the
self-annihilation rate is less than the expansion rate of the universe.
However it has recently been shown\cite{f:100} that when the gluino is
also light, other processes involving $\photino - R^0$ interconversion
and $R^0$ self-annihilation become important, and freezeout is delayed
to a lower temperature.  When the ratio $r =
\frac{m(R^0)}{m_{\photino}}$ is in the range 1.2 - 2.2, the relic
photino abundance can account for the dark matter of the
universe\cite{f:100}.  Only when $r \gsi 2.2$ would the photino relic
density be too large.  If that were the case, $R$-parity would have to
be violated, so that the photino is not absolutely stable, in order
for this scenario to be viable.

Let us turn now to discovery strategies for the $R^0$ and $S^0$.
With improvements in the determination of the MSSM' parameters by
careful study of the renormalization group and ew symmetry breaking
constraints\footnote{GRF and C. Kolda, in preparation.}, and use of
exact 1-loop radiative corrections\cite{pierce_papa} rather than the
approximate ones of ref. \cite{f:96}, it should be possible to rather
precisely predict the photino and gluino masses.  Lattice gauge
calcuations without quenched approximation could give the $R^0$
mass well enough to test whether the ratio $r =
\frac{m(R^0)}{m_{\gluino}}$ can lie in the range 1.2-2.2 needed to
explain the dark matter\cite{f:100}.  The $R^0$ lifetime can be
estimated using lattice gauge theory, which should be much more
satisfactory than the model developed above, to compute the hadronic
matrix elements of the short distance operators. It will take some time
for all these things to be done, so that for the present we should
consider discovery strategies in the two cases: that it can be
discovered via its decay, or it is too long lived for that.

In ref. \cite{f:95} I discussed strategies for detecting or excluding
the existance of an $R^0$ with a lifetime so long that it only rarely
decays in the apparatus.  If instead the $R^0$ lifetime is in the
$\sim 10^{-7} -  5 \times 10^{-11}$s range, it should be possible to take
advantage of the several very high-intensity kaon beams and the rare
kaon decay and $\epsilon'/\epsilon$ experiments, to find evidence for
the $R^0$.  It is fortuitous that the kaon experiments often run with
and without regenerator, and the $K^0_L$ and $K^0_S$ lifetimes are
comparable to the lifetimes of interest for the $R^0$, allowing a
large portion of the relevant range of lifetimes to be probed.  The
beams for such experiments contain $R^0$'s, whose decays one wants to
observe.  While $R^0$ production crossections can be reliably computed
in perturbative QCD when the $R^0$'s are produced with $p_{\perp} \gsi
1$ GeV, high-luminosity neutral kaon beams are produced at low
$p_{\perp}$ so pQCD cannot be used to estimate the $R^0$ flux in the
beam.  In this situation, a conservative lower bound on the production
cross section could be the production cross section of $\bar{\Xi}$, at
least for $m(R^0) \sim 1.4$ GeV.

The momentum in the $R^0$ rest frame of a hadron, $h$, produced in the
two body decay $R^0 \rightarrow \photino +~ h$, is:
\eq
P_h = \frac{\sqrt{m_R^4 + m_{\photino}^4 + m_h^4 - 2 m_R^2
m_{\photino}^2 - 2 m_{\photino}^2 m_h^2 - 2 m_h^2 m_R^2}}{2 m_R}.
\label{P}
\en
This falls in the range 300-800 MeV when $h= \pi^0$, for the mass
ranges of interest: $1.2 ~{\rm GeV} < m_R < 1.7 ~{\rm GeV}$ and $ 0.2
{}~{\rm GeV} < m_{\photino} < 0.9 ~{\rm GeV}$.  Therefore, unless the
$R^0$ is in the extreme high end of its mass range and the photino is
in the low end of its estimated mass range, final states with more
than one hadron will be significantly suppressed by phase
space\footnote{For instance the final state
$\pi^+\pi^-\pi^+\pi^-\photino$ suggested by Carlson and Sher, while
certainly distinctive, has a very small branching ratio for
practically all the masses under consideration.}.  A particularly
interesting decay to consider is $R^0 \rightarrow \eta \photino$.\footnote{I
thank W. Willis for this suggestion.}  Since $m(\eta) = 547~ {\rm
MeV} > m(K^0) = 498$ MeV, there would be very little background
mimicking $\eta$'s in a precision $K$-decay experiment, so that
detecting $\eta$'s in the decay region of one of these experiments
would be strong circumstantial evidence for an $R^0$.  Since the $R^0$
is a flavor singlet and the $\photino$ is a definite superposition of
isosinglet and isovector, the relative strength of the $R^0
\rightarrow \pi^0 \photino$ and $R^0 \rightarrow \eta \photino$ matrix
elements is determined by Clebsches and is $9:1$.  Thus the branching
fraction of the $\eta \photino$ decay mode is about 10\%, in the most
favorable case that multibody decay modes, and phase space suppression
of the $\eta$ relative to the $\pi^0$, are unimportant.  If $\eta$'s
are detected, the Jacobian peak in the $\eta$ transverse momentum,
which occurs at $p_{\perp} \approx P_{\eta}$ defined in eq. (\ref{P})
above,  gives both a confirming signature of its origin, and provides
information on the $R^0$ and $\photino$ masses.  The main issue in
this search is clearly resolution.  For long lifetimes, efficiency and
precise kinematic reconstruction are the principle concerns.  The
sensitivity for short lifetimes is dependent on good resolution of the
longitudinal location of the decay, to make sure one is not seeing
$\eta$'s produced in the regenerator.  It seems that the generation of
kaon experiments which will be running in the next year or so will be
able to make this search\footnote{I. Manelli and S. Somalwar, private
communications.}.

\begin{figure}
\epsfxsize=\hsize
\epsffile{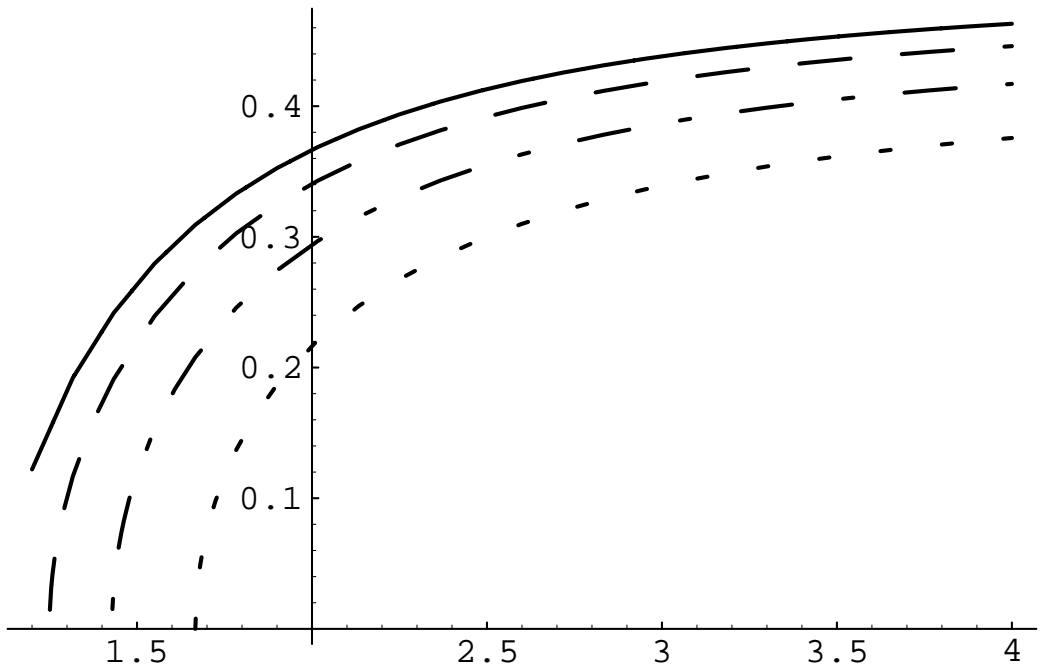}

\caption{$P_h$ in units of $m(R^0)$ as function of $r\equiv
\frac{m(R^0)}{m_{\photino}}$, for $\frac{m_h}{m(R^0)} = 0.1$ (solid),
0.2 (dashed), 0.3 (dot-dashed), and 0.4 (dotted).}
\label{P_h}
\end{figure}

Taking seriously the possibility that photinos account for the cold
dark matter of the universe leads us to be particularly interested in
the possibility that $r = \frac{m_R}{m_{\photino}}  \lsi 2.2$.  Fig.
\ref{P_h} shows how $P_h$ depends on $r$, from which we see that in the
region of interest the $R^0 \rightarrow \photino \eta$ decay is
considerably kinematically suppressed compared to $R^0 \rightarrow
\photino \pi^0$.  Thus it would be very attractive to also be able to
identify the latter reaction in the kaon decay experiments.  This is
much more demanding technically, but may be essential.  Given
measurements of $p_{\perp}^{max}$ in both $R^0 \rightarrow
\photino \eta$ and $R^0 \rightarrow \photino \pi^0$, one could
obtain at least a rough estimate of $r$, testing the dark matter
possibility.  This is because when $\frac{m_{\eta}^4 - m_{\pi}^4}{4
m_R^4} << 1$, eq. (\ref{P}) leads to $ P_{\pi}^2 - P_{\eta}^2 \approx
\frac{1}{2} (1 + \frac{1}{r^2}) (m_{\eta}^2 - m_{\pi}^2)$, allowing
$r$ to be extracted.

Light gluinos have many indirect consequences.  None of them are
presently capable of settling the question as to whether light
gluinos exist, since they all rely on understanding non-perturbative
aspects of QCD.  So far, our assessment of the inherent theoretical
uncertainty of QCD predictions is founded on the quality of the
agreement between models and data.  If the true theory is not QCD but,
say, QCD', this will lead to an underestimate of the uncertainty in
the predictions, since the models are tuned to agree with data
assuming the validity of standard QCD without gluons.  Nonetheless it
is interesting to recall some phenomena sensitive to differences
between QCD and QCD':
\begin{enumerate}

\item The running of $\alpha_s$ is different with and without light
gluinos.  In principle this can be investigated comparing
$\alpha_s$ determined at different scales, e.g., from deep-inelastic
scaling, and at the $\tau$, $\Upsilon$ and $Z^0$.  There are pitfalls,
but progress is being made. See \cite{f:95} for discussion and
references.  Less susceptible to modeling errors and very promising
when the fermionic determinant can be computed precisely enough, will
be to use consistency between data and lattice gauge calculations of
quarkonia spectra, with and without light gluinos\footnote{S. Shenker,
private communication.}.

\item  Jet production at FNAL and LEP is different with and without
light gluinos.  Since gluinos in this scenario are long enough lived
that they hadronize before decaying to a photino, they produce jets similar
to those produced by the other light, colored quanta: gluons and
quarks\cite{f:82}.  In $Z^0$ decay, only 4- and more- jet events are
modified; the magnitude of the expected change is smaller than the
uncertainty in the theoretical prediction.  Calculation of the 1-loop
corrections to the 4-jet amplitudes would allow the theoretical
uncertainty to be reduced sufficiently that data might be able to
discriminate between QCD and QCD'\cite{f:82}.  In $p \bar{p}$
collisions, there is a difference already in 1-jet cross
sections\cite{f:82}.  However absolute predictions are problematic
since they rely on structure functions which have been determined
assuming QCD, not QCD'.  This could be improved, but probably would
have large uncertainties.  More promising might be to search for
differences in the expected {\it relative} $n$-jets cross
sections\cite{f:82}.

\end{enumerate}

In summary, this paper has investigated the phenomenological
consequences of a restrictive form of SUSY breaking which is
theoretically attractive because it avoids cosmological problems
associated with gauge singlet fields and solves the SUSY CP problem.
Paramters of the theory were constrained by requiring correct
electroweak symmetry breaking, and masses and lifetimes of particles
were estimated.  The main conclusions, including some results obtained
in refs. \cite{f:95,f:96,f:100}, are the following:

\begin{itemize}

\item  Gluino mass is  100-600 MeV; photino mass is $\sim 100-1400$ MeV;
lightest chargino has a mass less than $m_W$; squark and slepton
masses will be of order 100-200 GeV\footnote{Conventional squark
limits do not apply when the gluino is light and long-enough-lived to
hadronize, as in this scenario.  Here, the relevant signal will be a
peak in the invariant mass distribution of a pair of jets, because
squarks decay via $S_q \rightarrow \gluino q$.  It is reasonable to
expect that the squarks associated with the $u,~d,~s,~c$ and $b$
quarks will be approximately degenerate, while the stop will be
significantly heavier.  Moreover the cross section for producing a
squark pair of each  flavor will be comparable within about a factor
of two with the cross section for producing a pair of heavy quarks of
that mass.  From the calculated production rates for $t \bar{t}$ it
is clear that there should be a substantial number of events
containing squark pairs at FNAL, up to quite high squark mass.  A
search for events in which {\it two} pairs of jets reconstruct to
the same invariant mass should be made.}; the $\mu$ parameter is $\lsi
100$ GeV and $tan \beta \lsi 2$.

\item The lightest R-hadron is probably the $R^0$, with mass $\sim 1.4 \pm
.3$ GeV and lifetime $\gsi {\rm few} \times 10^{-10}$s, possibly much
longer.  The decay mode $R^0 \rightarrow \eta \photino$ can have a
branching fraction of up to about 10\%.  Finding evidence of $\eta$
production in an intense neutral kaon beam, with missing $p_{\perp}$
having a jacobian peak characteristic of a two body decay, would be a
spectacular signal of this scenario.  For $m_{\photino} \gsi \frac{1}{2}
m(R^0)$, the decay $R^0 \rightarrow \pi^0 \photino$ is optimal,
albeit very challenging experimentally.  Some possibilities if the $R^0$
has too long a lifetime for detection through its decay are discussed
in ref. \cite{f:95}.

\item  The lightest color-singlet supersymmetric particle, the
photino, is an excellent cold dark matter candidate if $r \equiv
\frac{m(R^0)}{m_{\photino}}$ has a critical value in the range 1.2 -
2.2\cite{f:100}.  This is consistent with the masses expected in the
present scenario.  If the $R^0$ is discovered, determining its
lifetime and photino-production cross sections will allow a much more
precise computation of the critical value of $r$.  This would allow
confirmation or refutation of the proposal of ref. \cite{f:100}, that
relic photinos are responsible for the bulk of the missing matter of
the Universe.

\item There should be a flavor-singlet pseudoscalar with mass $\sim
1.4 \pm .4$ GeV, in addition to the mesons and glueballs of the conventional
QCD spectrum.  The MarkIII\cite{mark3} and DM2\cite{dm2} experiments
find evidence for two flavor singlet pseudoscalars and one vector in the
$\iota(1440)$ region, where only one pseudoscalar is expected in
ordinary QCD.  With its much greater statistical power, the Beijing
collider should be able to definitively determine the resonance
structure of this region.  Establishing the predicted extra pseudoscalar
would be a strong boost for the scenario advocated here.
\end{itemize}

{\bf Acknowledgements:}  I am deeply indebted to W. Willis and M.
Schwartz for discussions of search strategies, and to collaborators in
investigating various aspects of this problem: E. W. Kolb, C. Kolda,
M. Luty, A. Masiero and S. Thomas.




\end{document}